**Title**: Microbiomes and pathogen survival in crop residues, an ecotone between plant and soil

**Authors**: Lydie Kerdraon, Valérie Laval, Frédéric Suffert

**Address**: UMR BIOGER, INRA, AgroParisTech, Université Paris-Saclay, 78850 Thiverval-Grignon, France

frederic.suffert@inra.fr; tel +33 (0)1 30 8154 35; ORCID 0000-0001-6969-3878

**Abstract**

The negative contribution of crop residues as a source of inoculum for plant diseases is well established. However, microbial ecologists have long reported positive effects of residues on the stability of agrosystems and conservation tillage practices have become increasingly widespread. Most studies have suggested that large microbial communities should be taken into account in plant disease management, but we know little about their ecological interaction with pathogens in the crop residue compartment. This review focuses on microbiomes associated with residues within the context of other microbial habitats in cereal-producing agroecosystems such as phyllosphere or rhizosphere. We connect residue microbiome with the survival of residue-borne fungal plant pathogens, thus combining knowledge in microbial ecology and epidemiology, two disciplines still not sufficiently connected. We provide an overview of the impact of residues on cereal disease epidemics and how dynamic interactions between microbial communities of non-buried residues during their degradation, along with soil and multitude of abiotic factors, can contribute to innovative disease management strategies, including next-generation microbiome-based biocontrol strategies. Starting from the classical but still relevant view of crop residues as a source of pathogen inoculum, we first consider possibilities for limiting the amount of residues on the soil surface to reduce the pathogen pressure. We then describe residues as a transient half-plant/half-soil compartment constituting a key fully fledged microbial ecosystem: in other words, an ecotone which deserves special attention. We focus on microbial communities, the changes in these communities over time and the factors influencing them. Finally, we discuss how the interactions between the microbial communities and the pathogens present on residues could be used: identification of keystone taxa and beneficial assemblages, then preservation of these taxa by adapted agronomic practices or development of synthetic communities, rather than the introduction of a single exogenous biocontrol species designed as a "treatment product".

# Introduction

Crop residues, i.e. the decaying parts of the crop plant that is not harvested, are sometimes considered as waste. However, they are actually a source of essential environmental services necessary for the perpetuation of productive agrosystems (Smil, 1999). The conservation of crop residues in the field after harvest with reduced- or no-tillage practices allow the formation of soil organic carbon, improve soil structure, prevent erosion, filter and retain water, and reduce evaporation by maintaining a substantial part of crop residues on the soil surface (Derpsch *et al.*, 2010; Govaerts *et al.*, 2007). Yet, these practices, which are becoming increasingly common worldwide (Awada *et al.*, 2014; de Freitas and Landers, 2014; Kertész and Madarász, 2014), increase the risk of "residue-borne" or "stubble-borne" disease epidemics, due to the presence of crop residues acting as a source of primary inoculum (Bailey, 1996; Bailey and Lazarovits, 2003; Bockus and Shroyer, 1998). Indeed, several leaf-, stem-, and head-infecting microorganisms are known to survive on crop residues between cropping seasons (Bockus and Shroyer, 1998; Cook *et al.*, 1978). Residue conservation tends to increase the risk of epidemics for many foliar diseases, especially on cereals (Bailey, 1996; Bailey and Lazarovits, 2003; Bockus and Shroyer, 1998). For instance, wheat pathogens, such as *Pyrenophora tritici repentis* (Adee and Pfender, 1989), *Ocumimacula yallundae* (Vero and Murray, 2016), and *Zymoseptoria tritici* (Suffert and Sache, 2011), have been shown to be more likely to infect the subsequent crop if wheat residues are left in the field after harvest. The survival of such pathogens usually results from the formation of fruiting bodies arising from sexual reproduction on crop residues. These fruiting bodies, once mature, release spores for several months or even years, which can be spread over long distances and thus contaminate subsequent crops. The current review is focused on wheat residues, as cereal pathogens are agronomically important and well documented in the literature, but the ideas could be applicable to other crop/pathogen associations.

Epidemiological considerations have expanded to include new microbial ecological concepts, such as the "pathobiome", defined as the pathogen and the cohort of associated microorganisms likely to influence the persistence, transmission and evolution of the pathogen (Vayssier-Taussat *et al.*, 2014). This concept encompasses the "microbiota", defined as the assemblage of microorganisms present in a given environment, whereas the "microbiome" is defined as the entire habitat, including the microorganisms, their genomes, and the surrounding environmental conditions (Marchesi and Ravel, 2015) as a distinct additional node influenced by the three components of the classic epidemiological triangle (Suppl. Figure 1; Paulitz and Matta, 1999; Foxman and Rosenthal, 2013; Hanson and Weinstock, 2016; Legrand *et al.*, 2017). The "pathogen" is finally repositioned into a broader community context. It is considered as one of or many taxa responsible for disease that interacts positively or negatively with other taxa present in the same ecological niche. Disease expression seems to be the result of an imbalance between a potentially pathogenic species and the rest of the microbial community on host tissues, rather than simply a consequence of just the presence of the pathogenic species (Vayssier-Taussat *et al.*, 2014). The categorization of a species is rendered more complex by the fact that microorganisms reported to act as crop pathogens or endophytes can also develop without symptom development, or as saprotrophs in the soil and plant residues. Microbial assemblages harbour "keystone taxa" (Banerjee *et al*., 2018), that exert individually or in a guild, i.e. group of species that exploit the same resources, a considerable influence on microbiome structure and can drive community composition and function irrespective of their abundance across space and time. Thus, agronomic practices can theoretically cause a dramatic shift in microbiome structure and functioning by removing these keystone taxa, but can also reduce pathogen propagation and plant infection by beneficial microorganisms. The key questions are: which microorganisms should be considered as beneficial? Are these beneficial microorganisms keystone taxa within the community? Which management practices could preserve or promote these keystone taxa and enhance their positive activity against residue-borne diseases?

A large number of studies focused on plant and soil microbiomes, but the microbial communities specific to the crop residues left on the soil surface have been little investigated. The few studies concerning crop residues conducted to date have focused on the impact of buried residues on soil microbial communities, rather than on the residue microbial communities. In addition, they were performed in microcosms, with sterilized residues (Bastian *et al.*, 2009; Cookson *et al.*, 1998; Nicolardot *et al.*, 2007). Such study conditions are not optimal because they greatly decrease the complexity of the residue microbiome.

This thought-provoking review situates the microbiome associated with crop residues within the context of other microbial habitats in cereal-producing agroecosystems. We start from the classical but still relevant view of crop residues as a source of pathogen inoculum, highlighting the limits of their quantitative management. We then go on to better define crop residues as an ecological "compartment" that has strong spatial and temporal relationships with plant and soil compartments, firstly based on a "static" representation, and then on more dynamic processes and interactions. We pass from the notion of substrate to pathobiome by focusing on the microbial communities present on residues, changes in these communities over time and the factors likely to influence them. Finally, we reflect on the possible uses of the communities present on the residues and possible interactions with pathogens. We considered the possibility of identifying potential biocontrol agents from the cultivable part of these communities by microbiome-based strategies.

# 1. Impact of crop residues on the development of fungal disease epidemics: cereal cropping systems as case study

Tillage, defined as the mechanical manipulation of soil and plant residues for seedbed preparation (Reicosky and Allmaras, 2003), has been associated with agriculture for several millennia. Deep tillage systems has several benefits, including weed control, greater yields and root length densities for some crop species (Varsa *et al.*, 1997), but tillage also has negative consequences on the soil: disruption of structure, erosion (Borrelli *et al.*, 2017), and reduction in organic carbon depending on factors such as depth of measurement, soil type, and tillage method (Haddaway *et al.*, 2016). No-tillage has repeatedly been shown to have beneficial effects on soil preservation. The retention of crop residues at the soil surface prevents water erosion, by reducing the direct impact of raindrops (Chambers *et al.*, 2000; Hobbs, 2007), reducing runoff velocity and giving water longer to infiltrate (Pimentel *et al.*, 1995). It also prevents wind erosion by protecting the soil and enhancing the soil's physical, chemical, and biological properties (Kassam *et al.*, 2015; Verhulst *et al.*, 2010). No-tillage has another positive impact, by decreasing the emissions from farming activities through the reduction of mechanical operations (Govaerts *et al.*, 2007). Finally, conservation agriculture and no-tillage practices, and the use of permanent soil cover and rotations (Hobbs, 2007) have steadily increased from 2.8 million ha in 1973/1974, to 110.8 million ha in 2007/2008 in the world (Derpsch *et al.*, 2010).

The retention of residues at the soil surface can also have a negative effect: the promotion of so-called "residue-borne" diseases. These crop diseases, many of which are foliar, are mainly caused by fungal pathogens that can overwinter on residues by carrying out a specific part of their life cycles leading to the production of primary inoculum (Dyer *et al.*, 1996; Leplat *et al.*, 2013; Shaw and Royle, 1989; Suffert and Sache, 2011; Vero and Murray, 2016). Unburied crop residues lying on the soil surface in low-tillage systems can be seen as a "brown bridge" of dead plant material that can harbor multiple pathogenic, saprotrophic or endophytic species (Thompson *et al.*, 2015). Crop residue management is a particularly important issue as residues can have a major impact as a recurrent source of inoculum over long periods, often exceeding the interepidemic period during the intercropping season when host plants are not present in the agrosystem. For instance, spores of *Fusarium* species, a pathogen of cereals such as wheat and maize, could be released for up to 3 years after harvest from maize residues (Pereyra *et al.*, 2004).

The survival of plant pathogens on crop residues was shown to be inversely correlated with the amount of residues buried or exported after harvest (Bockus and Claassen, 1992; Carignano *et al.*, 2008; Dill-Macky and Jones, 2000; Guo *et al.*, 2005; Jørgensen and Olsen, 2007) and with the degree of residue degradation (Gosende *et al.*, 2003; Hershman and Perkins, 1995; Leplat *et al.*, 2013; Marcroft *et al.*, 2003; Pereyra *et al.*, 2004; Pereyra and Dill-Macky, 2008). However, it is difficult to generalize the quantitative impact of residues as an effective source of inoculum (e.g. Morais *et al.*, 2016), because the nature of the survival structures depends on the biology of the pathogen and the environmental conditions.

Most disease management strategies target the epidemic period of the disease, although interepidemic period is crucial for pathogen survival, as highlighted above. Decreasing the presence of a pathogen during the interepidemic period could theoretically limit disease development in the next crop and even over the next few years. A correlation between the amount of primary inoculum at the end of the growing season and disease severity during the next season has been established for *Leptosphaeria maculans* pathogen of oilseed rape (Lô-Pelzer *et al.*, 2009) and *Pyrenophora tritici repentis* (Adee and Pfender, 1989; Bockus and

Claassen, 1992), *Paragonospora nodorum* (Mehra *et al.*, 2015), and *Zymoseptoria tritici* (Suffert *et al.*, 2018) pathogens of wheat.

A number of studies have demonstrated the value of managing the primary inoculum to limit disease severity during the epidemic period (Adee and Pfender, 1989; Filho *et al.*, 2016), but in several cases the amount of primary inoculum is not a limiting factor for the development of epidemics (Alabouvette *et al.*, 2006). This is the case, for example, for polycyclic diseases caused by pathogens capable of multiple infection cycles with inoculum increases throughout the growing season (Burdon and Laine, 2019).

2. **Crop residues, a shifting compartment hosting microbial communities interacting with plant and soil**

    2.1. **The residues, ecotone in the interface between the plant and soil: definitions and concepts**

Buried and non-buried crop residues should be distinguished according to their location: above- and below-ground, respectively (Figure 1). This distinction makes perfect sense in terms of the epidemiology of plant diseases, as buried residues are no longer a source of inoculum for airborne diseases as highlighted above. The term "residuesphere" identifies the microhabitat consisting of all crop residues, whether buried or non-buried. However, this term has occasionally been used as a synonym of "detritusphere" (Magid *et al.*, 2006; Sengeløv *et al.*, 2000), which is defined as the soil adjacent to plant residues (Marschner *et al.*, 2011; Pascault *et al.*, 2010a; Poll *et al.*, 2008). The detritusphere is considered to be the part of the soil immediately affected by residue degradation, and is generally assumed to include the first 6 mm (Bastian *et al.*, 2009; Nicolardot *et al.*, 2007) or 10 mm (Magid *et al.*, 2006) of soil surrounding the residues at all times in stratified experiments. This top layer of the soil is very thin but has high levels of microbial activity (Kuzyakov and Blagodatskaya, 2015). The detritusphere has been also defined ambiguously by some authors as the layer of soil including the litter and the adjacent soil influenced by the litter (Gaillard *et al.*, 1999; Ingwersen *et al.*, 2008). An experimental comparison of the various soil zones (residues, detritusphere, and bulk soil) indicated that the bacterial and fungal communities are specific to a residue type in the detritusphere and to the location of residues (Nicolardot *et al.*, 2007). Residue degradation has been shown to induce a particular genetic structure of the microbial community with a gradient from residue to bulk soil. Based on these findings, it was concluded that the residues, detritusphere and bulk soil corresponded to different trophic and functional niches for microorganisms.

Residues should be considered as a distinct microbial substrate (Bastian *et al.*, 2009), characterized by the plant from which they originate, and by their degradation stage, implying chemical and physical changes dependent on their position relative to the soil surface. Residues originate from the plant (temporal link; see part 3.2), are in close contact with the soil (spatial link; see part 3.3) and decay over the following cropping season, at rates dependent on plant species, cropping practices (Hadas *et al.*, 2004), and year (climate effect). Residues are definitely not a specific "static" compartment and should be viewed as "transient" (Figure 2).

### 2.2. Microbial communities of residues are inherited from the plant

The plant hosting the "phytobiome", which consists of the plant, communities of organisms – micro- and macro-organisms – present in and on plants, includes the phyllosphere, rhizosphere and endophytic compartments (Figure 1). These three compartments make a crucial contribution to the crop residues. Microorganisms are adapted to particular ecological niches and physiological conditions (Grudzinska-Sterno *et al.*, 2016; Larran *et al.*, 2007; Vorholt, 2012) and are driven by numerous biotic and abiotic factors (e.g. temperature, humidity, light; Carvalho and Castillo, 2018), including the plant itself, whether considered at the species or genotype level (Bodenhausen *et al.*, 2014; Wagner *et al.*, 2016). The effect of the plant is therefore mainly due to two aspects: (i) it is already colonized by microorganisms that can remain on the residues (e.g. hemibiotrophic pathogens) and (ii) it has different biochemical compositions, which can affect the rate of degradation and the changes in the chemical and physical properties of the residues. These two aspects are closely linked, because the microorganisms present on plants are partly dependent on the biochemistry of the plant (species, genotype) and because these microorganisms can have a reciprocal impact on the plant properties, even before any degradation has occurred (changes in C/N ratio, production of defense compounds by plants in response to colonization by microorganisms, etc.). Both aspects must be taken into account when trying to understand how new microorganisms colonize residues and how their microbiota changes over time in the agrosystem.

Green tissues of plants are naturally infected with endophytic species. Biocontrol engineering practices based on artificial inoculation with such endophytic species are not yet largely used with exogenous fungal or bacterial strains during a cropping season, probably because of the complexity of the microbial interactions involved and their alteration by variations in agro-environmental conditions in field conditions (De Silva *et al.*, 2019). Nevertheless, it is established that endophytic species naturally present in cultivated plants can limit the development of pathogens during the cropping season (Stone *et al.*, 2018). It is therefore likely that their action would continue on the residues during the intercropping season.

The microbial communities present on the residues at the beginning of their degradation depend on the plant (species, genotype, organ) and on a pool of other organisms that differ according to the events in the plant's life and the environment in which it was grown (biotic and abiotic stresses). Even after the residues come into contact with the soil, the influence of the plant species on residue colonization is evident (Nicolardot *et al.*, 2007; Kerdraon *et al.*, 2019a). Plant genotype also determines the structure of the microbial communities of the whole phyllosphere, particularly in cases of resistance to certain pathogens, but also for epiphytes and endophytes in the case of wheat (Sapkota *et al.*, 2015; Karlsson *et al*., 2017). Comby *et al.* (2017) showed that the temporal variation of wheat microbial communities in natural conditions was driven by a succession of plant pathogens. The communities present on plants also depend on the organ considered. This is particularly true for aerial organs and roots, because of the differences between habitats in terms of nutrients and exposure. However, several species, such as *Cladosporium* sp. and *Michrodochium nivale,* can colonize all parts of the plant (Gdanetz and Trail, 2017; Grudzinska-Sterno *et al.*, 2016). Moreover, the age of each organ has been shown to determine the communities present (Wagner *et al.*, 2016). In some cases, the proportion of pathogenic fungi increases with plant development (Grudzinska-Sterno *et al.*, 2016). Crop rotation does not appear to affect the communities associated with leaves, but does seem to affect root-associated communities, as demonstrated in a wheat-pea rotation (Granzow *et al.*, 2017). It can be concluded from these various examples that pathogenic and non-pathogenic microorganisms present on plants remain on the residues and influence the subsequent dynamics of colonization by other microorganisms.

Microbial abundance and composition of the phyllosphere changes over time, which drives colonization of residues by specific microorganisms over the cropping season (e.g. Peñuelas *et al*., 2014). The biochemical composition of crop residues, which depends on the plant species from which they are derived, is one of the factors determining the structure and diversity of bacterial communities (Baumann *et al.*, 2009; Pascault *et al.*, 2010b). Residue degradation has been shown to depend on the complex chemical composition of the residues, taking into account the C/N ratio, and the nitrogen and lignin contents of the plant (Kriaučiuniene *et al.*, 2012). The colonizing microbial communities differ between plant species, on identical soils. This has been established, for example, by comparing the degradation dynamic of different crop residues (soybean, corn, wheat, oilseed rape, alfalfa; Broder and Wagner, 1988; Pascault *et al.*, 2010a). Differences between "recalcitrant" and "easily degradable" residues that underwent faster changes, were related to differences in the microbial communities present: those rich in organic substrates consisted mostly of copiotrophic genera whereas those from an environment poor in organic substrates consisted mostly of oligotrophic genera. Furthermore, fungi are considered as the main decomposers of recalcitrant compounds, whereas bacteria break down simple substrates (Boer *et al.*, 2005).

The gradual degradation of crop residues induce changes over time in the structure of microbial communities, and these microbial communities have a reciprocal effect, inducing biochemical transformations in the crop residues hosting them. A change in the bacterial and fungal communities present on wheat residues has been observed between the early stages of degradation (14 and 56 days after the incorporation of residues into the soil) and later stages (56 to 168 days) (Bastian *et al.*, 2009). This change was interpreted as a modification in the balance between copiotrophic and oligotrophic organisms. Residue communities are also influenced by cultivation techniques: residues are degraded differently in aerobic and anaerobic conditions (Cookson *et al.*, 1998). This results in different mobilized communities, including different bacterial and fungal cellulolytic species (Boer *et al.*, 2005), as degradation processes are not the same (Nicolardot *et al.*, 2007). The functional composition of the microbial communities (cellulolytic vs. lignolytic) depends on the physicochemical properties of the soil, such as pH.

## 2.3. Impact of the soil on changes in the microbial communities of non-buried residues during their degradation

Separating the "plant inheritance" effects (see part 2.2) from the "soil" effects is difficult, especially when considering degradation processes. It is easier for species whose origin can be determined with certainty because of specificities of their life cycle. For example, certain strictly biotrophic fungi such as rusts (Puccinia) or mildew (Erysiphe) are typically found in the phyllosphere, whereas strictly telluric bacteria or arbuscular mycorrhizal fungi will be observed in the bulk soil. The addition of crop residues to soil results in considerable heterogeneity in soil microbial community diversity, where three distinct zones have been identified: (i) the residues themselves, (ii) the detritusphere (the soil zone in close contact with the residues, see section 2.1), and (iii) the bulk soil (the distant soil zone unaffected by residue decomposition) (Nicolardot *et al.* (2007). The greatest microbial diversity dynamics was observed on the residues, suggesting that the distinction between these compartments should be not only "static" (Figure 1) but also "dynamic" as microbial interactions change over the residue degradation (Figure 2) (Nicolardot *et al.* (2007).

There are still significant gaps in our knowledge of functional ecology and diversity of microbial communities present on non-buried residues and their interaction with bulk soil. The residue mulch is suspected to be mainly decomposed by the fungal communities since soil under no tillage contains greater proportions of fungi to bacteria than those under conventional tillage, while bacteria are generally considered to be the predominant decomposers of incorporated crop residues under conventional tillage (Navarro-Noya *et al.*, 2014; Anderson *et al.*, 2017 ; Hellequin *et al.*, 2018). Some studies performed under controlled laboratory conditions help characterize the impact of crop residue inputs on the diversity of soil microbial communities (e.g. Bastian *et al.*, 2009; Nicolardot *et al.*, 2007), but the converse relationship has been rarely investigated. Finally, it is difficult to know whether the results of such studies, beyond the methodological aspects (e.g. metabarcoding), can be extrapolated to understand the impact of bulk soil on changes in microbial communities hosted by residues during their degradation.

Evidence for a major impact of soil compartment on the dynamics of microbial communities can only be indirect, such as the comparison of communities, and partial by characterization of ecological functions. For instance, the importance of soil microorganisms in the mineralization of plant residues is well established (Wardle *et al.*, 2004; Shahbaz *et al.*, 2017), but the microbial colonization of residues left on the surface, for example in autumn, 3-4 months after harvest when the production of pathogen spores released from residues of infected wheat peaks, is not well documented.

The degree of contact between crop residues and the bulk soil, which is determined by the amount of residues left on the soil surface and the intensity of incorporation in the soil, affects degradation dynamics under both natural and experimental conditions (Henriksen and Breland, 2002). Concretely, poor residue-soil contact reduces the decomposition of structural plant constituents by delaying colonization with microorganisms degrading cellulose and hemicellulose. Several studies focused on the effect of the location of residues (incorporated vs. left on the soil surface) and of organic matter on their subsequent degradation, both in large-scale field experiments (e.g. Pascault *et al.*, 2010a; 2013; Tardy *et al.*, 2015; Guo *et al.*, 2016; Hartman *et al.*, 2018; Xia *et al.*, 2019) and controlled conditions experiments (e.g. Bastian *et al.*, 2009; Nicolardot *et al.*, 2007). The aforementioned studies investigated the effect of different management options of residues on soil microbial communities, but very few have looked at the interactions between residue-borne pathogens and the microbiota driving residue degradation. This gap need to be addressed in the future. The ecological richness of the residue compartment at the interface between the plant and soil communities suggests potentially interesting prospects for next-generation biocontrol strategies. Indeed, residues can be considered as both a source and a target for biocontrol: a source, because the probability to identify interacting pathogen and beneficial species is higher than in the plant or soil compartment; and a target because the residues bearing plant pathogens could be treated in order to limit the amount of primary inoculum for several important diseases.

## 3. Residue microbiota and the interface between the plant and soil communities: prospects for microbiome-based biocontrol solutions

This review highlights that crop residues are transient half-plant/half-soil compartment, which constitute a key fully fledged microbial ecosystem. This residue microbial system represents an ecotone, as a boundary compartment between two biomes, the plant and the soil. The residue microbiota should be taken into account in the management of residue-borne diseases. It may be possible to identify guilds of beneficial microorganisms naturally present on residues, which could then be preserved, or even selected, characterized and used as biological control agents against the pathogens that complete their life cycle on residues.

Certain residue-borne fungal diseases, such as spot blotch and common root rot on wheat and barley caused by *Cochliobolus sativus* (Bailey and Lazarovits, 2003), tan spot on wheat caused by *P. tritici-repentis* (Bockus, 1998; Bockus and Claassen, 1992), and blackleg on canola caused by *Leptosphaeria maculans* (Guo *et al.*, 2005) can be reduced by decreasing the amount of residues on the soil surface. For most other polycyclic plant diseases caused by residue-borne pathogens, the management of crop residues at the field scale may be ineffective for significantly reducing the final disease severity and yield losses in a sustained manner. Such a decrease in pathogen load could only be achieved by limiting the primary inoculum over a larger scale, i.e. in all the fields located at a distance less than the approximate dispersal capacity of this inoculum. This would require combining knowledge of the rate at which spores of pathogens strongly decline with distance from inoculum sources. This review suggests that another complementary approach based on biocontrol strategy exploiting residue microbiota should be considered.

There are currently no examples or studies supporting the strategy of exploiting residue microbiota as biocontrol agents that have led to efficient and practical solutions applied in cereal cropping systems at a large scale. However, plant microbiomes have recently been suggested as the key to future targeted and predictive biocontrol approaches (Berg et al., 2017). For instance, some microorganisms commonly present in the phyllosphere (*Paecylomyces lilacinus*, *Fusarium moniliforme* var. *anthophilum*, *Epicoccum nigrum*, *Bacillus* sp., *Cryptococcus* sp. and *Nigrospora sphaerica*) have been shown to affect the germination of *Z. tritici* spores (Perello *et al.*, 2002). Some members of the microbial community may also have an impact on plant resistance to certain diseases (Ritpitakphong *et al.*, 2016). A number of effects can be targeted on residues, including increasing the rate of residue degradation by cellulose- or lignin-degrading microorganisms (e.g. Dinis *et al*., 2009), shortening indirectly the survival of certain organisms and promoting interactions affecting the saprotrophic development of pathogens, and thus the limitation of primary inoculum production. Some studies have reported beneficial effects of cropping practices, such as decreases in *F. graminearum* survival due to an increase in the population of microbial soil antagonists induced by the addition of green manure to the soil (e.g. Bailey and Lazarovits, 2003; Wiggins, 2003; Perez *et al.*, 2008; Zhang *et al.*, 2017). A recent study combining metabarcoding and co-occurrence network analysis profiled cereal microbial communities presents in maize residues and their potential interactions with the different pathogenic *Fusarium* species (Cobo-Díaz *et al.*, 2019). The microbial communities present in the maize residues may represent important taxa that could lead to biocontrol strategies against Fusarium Head Blight.

Given the high diversity of microorganisms on residues, various modes of action could be used to increase biocontrol efficiency (Alabouvette *et al.*, 2006; Guetsky *et al.*, 2001). Diverse modes of action were described for various organisms present on the residues. To our knowledge, only one study has reported biocontrol tests performed on crop residues (chickpea) infected with a residue-borne pathogen (*Didymella rabiei*). For example, *A. pullulans* can grow faster than *D. rabiei,* thereby limiting its propagation by competition; and *Clonostachys rosea*, which has mycoparasitic capacity, can decrease or even totally abolish the sexual and asexual reproduction of *D. rabiei* (Dugan *et al.*, 2005). *Microsphaeropsis* sp., which is also known to have mycoparasitic capacity (Benyagoub *et al.*, 1998), has been shown to affect the production of *Fusarium graminearum* spores on wheat and maize residues (Bujold *et al.*, 2001; Legrand *et al.*, 2017). Bastian *et al.* (2009) highlighted the colonization of sterile residues deposited back on the soil by bacteria such as *Pseudomonas fluorescens*, *Pseudomonas aurantiaca* and *Pseudomonas putida* and fungi such as *Chaetomium globosum*, which have been described as potential biocontrol agents (Clarkson and Lucas, 1993; Cordero *et al.*, 2014; Flaishman *et al.*, 1996; Kildea *et al.*, 2008; Larran *et al.*, 2016; Perello *et al.*, 2002; Pfender *et al.*, 1993; Ramarathnam and Dilantha Fernando, 2006).

Macroinvertebrates can also have an impact on the residue microbiome. Some empirical studies demonstrated that earthworms (e.g. *Lumbricus terrestris*; Wolfarth *et al*., 2011) can reduce the *Fusarium culmorum* biomass and deoxynivalenol concentration in wheat straw left on the soil surface. Amongst detritivorous species, collembolans (e.g. *Folsomia candida*) and nematodes (e.g. *Aphelenchoides saprophilus*) take also an important role in the control of phytopathogenic and toxinogenic fungi surviving on crop residues (Wolfarth *et al*., 2013; 2016): these organisms can be viewed as another potential driver to compensate negative consequences of conservation tillage.

**Perspectives**

Would it be possible to reduce the presence and activity of a pathogen in the microbial community of crop residues, through interactions with other species: competition for resources, antagonism, or parasitism? If this is, indeed, possible, it would not be performed by classical biocontrol consisting of the introduction of an exogenous microorganism, as such approaches have not always translated well under field conditions. The limitation of pathogen impact would be probably through the exploitation of functional, agro-ecological relationships to identify a guild of microorganisms consisting of one or several keystone taxa that compose the "beneficial fraction" of the community and then to promote these microorganisms. The challenge concerns both the identification of the microorganisms and their promotion, but to make advances, the factors favoring microorganisms must also be identified. This issue is a huge bottleneck to success and should be of growing importance for both academic and operational research working towards the development of microbiome-based biocontrol solutions. An accurate descriptive approach and the characterization of interactions within the residue microbiota are required. Next-generation sequencing is a promising technology for this approach (Toju *et al*., 2018). It provides access to the diversity of "non-culturable" microbes, facilitating the discovery of new species (Lagier *et al*., 2016) and more detailed community description (e.g. Kerdraon *et al*., 2019a). Co-occurence network analysis can help to identify potentially important endemic species with biocontrol activity by obtaining a better understanding of ecological processes (e.g. Kerdraon *et al*., 2019b), but a 'co-occurrence species' per se does not mean 'real activity'. Moreover, keystone taxa identified by this way are probably not the only species that may be playing a critical role within a microbial community and the strategy of co-occurence network analysis could be expanded to characterize the activity of exogeneous species introduced as biocontrol agents and their impact on the whole endogenous community.

Some of the next-generation sequencing techniques available could be used to characterize the diversity of the microbial communities associated with the pathogen throughout their life cycle during both the epidemic and interepidemic periods, even if these periods are cryptic. We must be aware that the epidemic period is probably not the only period to target: the objective is not to find a species that can replace a fungicide applied during crop growth, but also to understand how species, from single taxa to more complex microbial assemblage, can reduce the amount of primary inoculum during the interepidemic period. Non-culturable microorganisms clearly cannot be used for biocontrol methods involving the introduction of exogenous species, but we need to know more about their activity in natural conditions. It would then be possible to test, *in vitro* or *in planta,* the culturable species isolated from residues and identified as potential biocontrol agents, based on integrative strategies focusing on plants during their development (e.g. Gdanetz and Trail, 2017). To date, practical biocontrol solutions were usually based on single beneficial strains. This through-provoking review suggests that such solutions may be improved by developing microbiome-based strategies based on more complex microbial assemblage such as "synthetic communities", that can reintroduce natural complexity (Herrera Paredes *et al*., 2018; Sergaki *et al*., 2018). Tsolakidou *et al*. (2018) recently proposed that microbial synthetic communities can be used as compost inoculants to produce composts with desired characteristics. The creation of synthetic ecosystems using microbial mixtures can lead to promising solutions for sustainable agricultural practices. In that direction, further study is needed to understand how synthetic communities persevere in crop residues and what are the additive traits of a community compared to single species. Residues considered as an ecotone in the interface between the plant and soil are a compartment particularly rich in microorganisms, which is ideal not only for identifying assemblages of interest but also as pathogen inoculum source being itself the target of biocontrol synthetic communities.

**Acknowledgments**

This review was supported by a 2015-2019 grant from the European Union Horizon Framework 2020 Program (Grant Agreement no. 634179, EMPHASIS project) and a 2014-2016 grant overseen by the French National Research Agency (ANR) as part of the "Investissements d'Avenir" programme (LabEx BASC; ANR-11-LABX-0034; API-SMAL flagship project). We thank Julie Sappa for her help correcting our English. We also sincerely thank Dr. Carolyn Young and the five anonymous reviewers for their insightful comments of the manuscript and their enthusiasm for the subject.

**Figure captions**

**Figure 1**. Positioning of the crop residues in relation to the other compartments constituting the different ecological niches of an agrosystem.

**Figure 2**. Representation of the most significant dynamic relationships – flow of microorganisms (orange arrows), biotic interactions (black arrows), abiotic or cropping effects (blue arrows) – between the crop residue, plant and soil compartments in an agrosystem.

**Suppl. Figure 1**. Disease triangle, serving as a conceptual model, presenting the factors that interact to cause a plant disease epidemic, completed by microbiota as a "fourth node" (adapted from Paulitz and Matta (1999) and Legrand *et al*. (2017)).

Figure 1

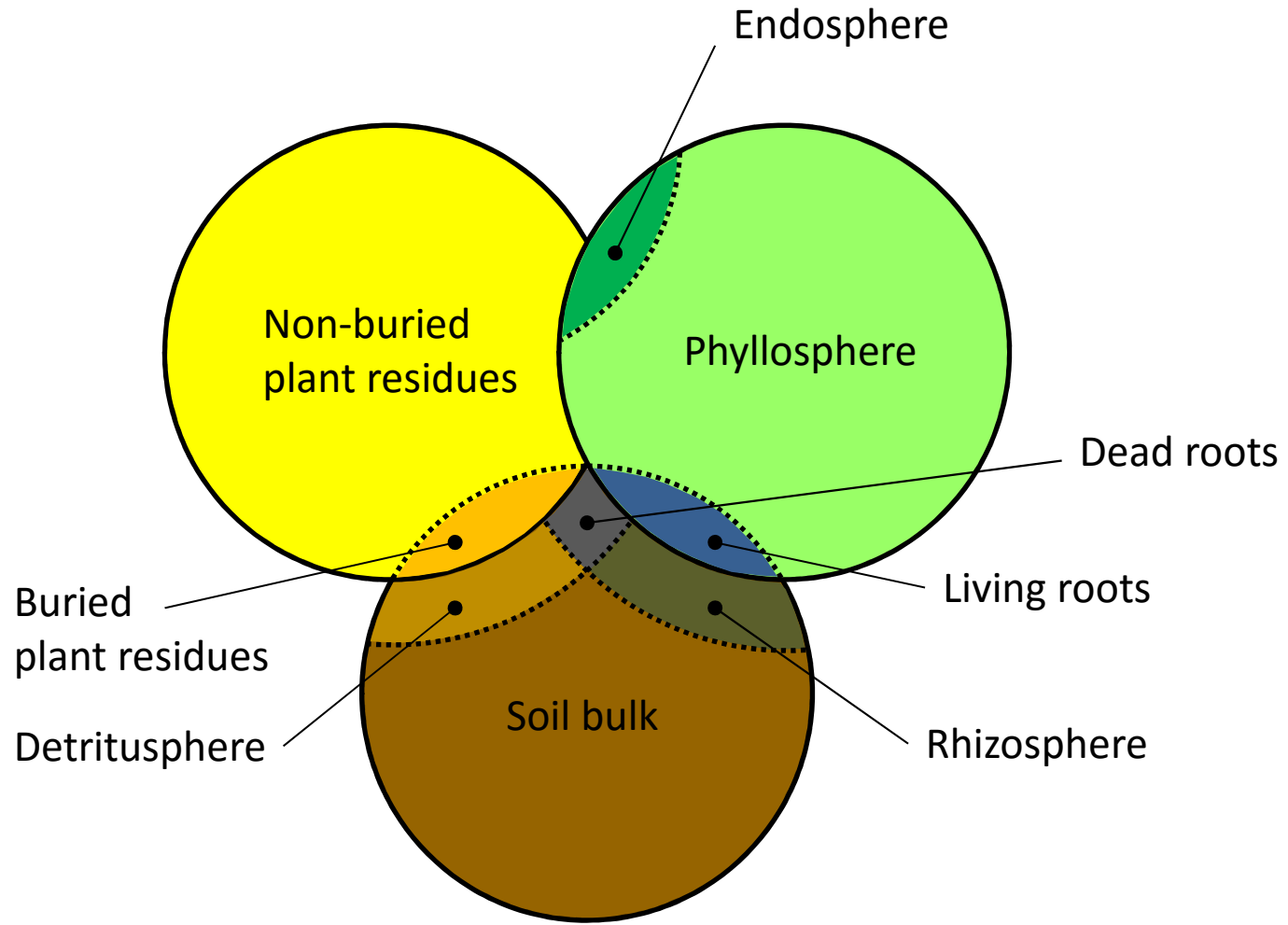

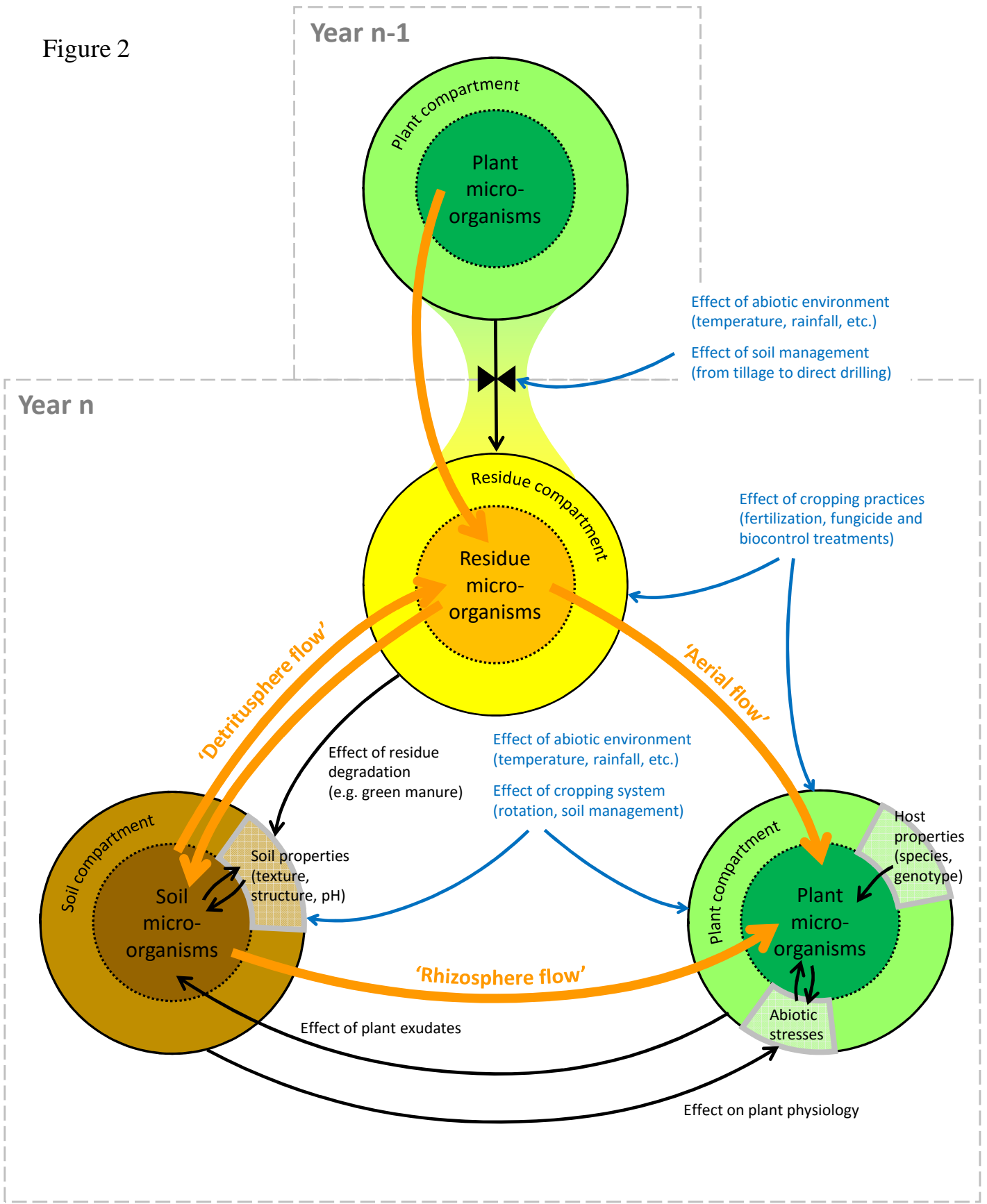

Supplementary Figure 1

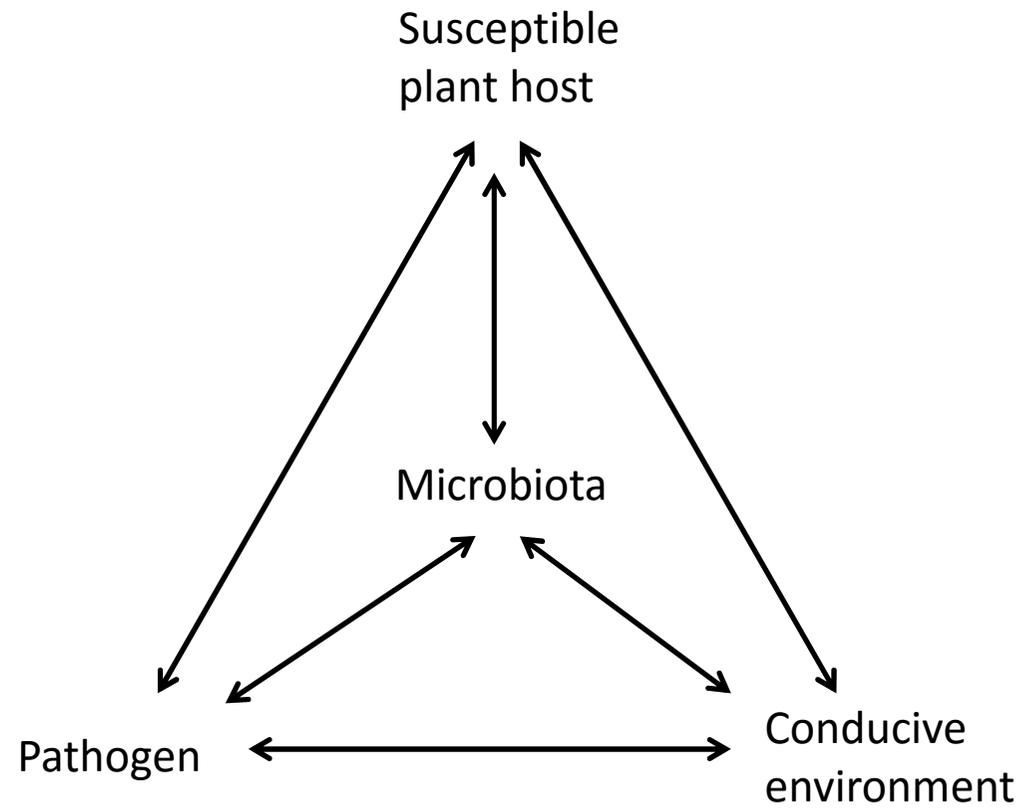